\documentclass[twoside,10pt]{amsart}

\usepackage{amsmath,graphicx,amssymb,fancyhdr,amsthm,enumerate,textcomp} 
\usepackage[usenames]{color}

\usepackage{hyperref}
\usepackage{float}
\usepackage{subcaption}
\usepackage[english]{babel}
\usepackage[numbers]{natbib}

\newtheorem{thm}{Theorem}[section]

\newtheorem{lem}[thm]{Lemma}

\theoremstyle{definition}
	
\theoremstyle{remark}

\def\beq{\begin{eqnarray}}
\def\eeq{\end{eqnarray}}
\def\bem{\begin{matrix}}
\def\eem{\end{matrix}}

\def\bsp{\begin{split}}
\def\esp{\end{split}}

\newcommand{\be}{\begin{equation}}
\newcommand{\ee}{\end{equation}}

\def \hbm #1 {\mbox{\boldmath{$\hat m^{(#1)}$}}}

\def \bm #1 {\mbox{\boldmath{$m^{(#1)}$}}}
\def \BDM {\begin{displaymath}}
\def \EDM {\end{displaymath}}

\newcommand{\Drr}{D''}

\newcommand{\Hr}{H'}

\newcommand{\Izero}{I_{0}}

\begin{document}


\title{Scalar Polynomial Curvature Invariants in the Context of the Cartan-Karlhede Algorithm}


\author{{\large\textbf{D. A. Brooks$^{1}$,~D. D. McNutt$^{1,3}$, ~ J. P. Simard$^{2}$,~\\ and~ N. K. Musoke$^{1,4}$ }} \vspace{0.3cm} \\
$^{1}$Department of Mathematics and Statistics,\\ Dalhousie University, Halifax, Nova Scotia,\\ Canada B3H 3J5 \\ 
$^{2}$ Department of Mathematics and Statistics, \\ University of New Brunswick, Fredericton, New Brunswick \\ 
Canada, E3B 5A3 \\
$^{3}$ Faculty of Science and Technology,\\ 
                         University of Stavanger, 
                         N-4036 Stavanger, Norway \\
$^{4}$ University of Auckland, \\ Auckland,  New Zealand 1142 \\ 
}
\date{\today} \maketitle \pagestyle{fancy} \fancyhead{} \fancyhead[CE]{Brooks et al.} \fancyhead[LE,RO]{\thepage}
\fancyhead[CO]{Scalar Polynomial Curvature Invariants in the Context of the C.K. Algorithm} \fancyfoot{} 

\begin{abstract} 

We employ the Cartan-Karlhede algorithm in order to completely characterize the class of G{\"o}del-like spacetimes for three-dimensional gravity. By examining the permitted Segre types (or P-types) for the Ricci tensor we present the results of the Cartan-Karlhede algorithm for each subclass in terms of the algebraically independent Cartan invariants at each order. Using this smaller subset of Cartan invariants we express the scalar polynomial curvature invariants for the G{\"o}del-like spacetimes in terms of this subset of Cartan invariants and generate a minimal set of scalar polynomial curvature invariants that uniquely characterize metrics in the class of G{\"o}del-like spacetimes and identify the subclasses in terms of the P-types of the Ricci tensor.

\end{abstract} 

\maketitle

\begin{section}{Introduction}

The question of when two spacetimes  in general relativity can be transformed from one to another by a change of coordinates is known as the equivalence problem. This is an important issue for solutions to the Einstein field equations, as one would like to know when two solutions are related and hence describe the same gravitational field. It can also be difficult to determine whether an effect derived from the metric is due to the coordinates chosen or is of a real physical nature. The solution to the equivalence problem concerns itself with obtaining an invariant characterization of the local geometry of the spacetime, and provides a framework from which the answers to these and other questions can be obtained \cite{Sousa,MW2013}. In general, having an invariant description of a metric can be quite applicable and has far-reaching consequences \cite{Olver, MacCallum2015, gang}.

Mathematically, interest in the equivalence problem goes back to the time of Gauss \cite{Olver}. Christoffel was the first to investigate this problem for $n$-dimensional Riemannian manifolds admitting no symmetries, and his work implied that for $n=4$ the twentieth covariant derivative of the curvature tensor was required for complete classification \cite{Ehlr}. Cartan applied his method of moving frames in 1946 to the classification of Lie pseudo-groups, and extended this approach to metrics admitting symmetries. In general, this approach required comparing the components of the curvature tensor and its covariant derivatives up to $n(n+1)/2$-th order \cite{Olver, Cartan}. Sternberg completed Cartan's approach in the 1960's, and with the development of computer algebra systems Karlhede gave Cartan's approach a true algorithmic form and adapted it specifically for general relativity in 1980 \cite{Karl1, Karl2}. Algorithms using the algebraic classification of the irreducible parts of the curvature spinor and the Newman-Penrose spinor formalism have also been developed and implemented for various computer algebra suites \cite{Hall, AlievNutku, Pen1}.

To determine if two spacetimes are equivalent we must construct a set of invariants that are necessary and sufficient to characterize uniquely each spacetime. In practice, there are two major approaches used to generate invariants: the Cartan-Karlhede  (C.K.) algorithm and scalar polynomial invariants (SPIs). Implementing the C.K. algorithm for three-dimensional (3D) solutions requires knowledge of the coframe formalism and the effect of frame transformations on the Ricci tensor and its higher order derivatives. In comparison the computation of SPIs is straightforward, one merely takes the copies of the curvature tensor and its covariant derivatives, and contract the indices to produce scalars. This procedure will give the same result regardless of the choice of basis. Admittedly, the set of all SPIs, $\mathcal{I}$, is not sufficient to uniquely characterize all spacetimes. It has been shown that if a  spacetime is not uniquely characterized by $\mathcal{I}$, it is either locally homogeneous or it  belongs to the class of degenerate Kundt metrics \cite{CSI1, CSI2, CSI3, Kundt2009}.  

For the class of spacetimes uniquely characterized by $\mathcal{I}$, there are several unanswered questions relating to the minimal number of SPIs to compute, and the highest order covariant derivative of the curvature tensor needed to completely classify a spacetime. In four dimensions (4D) considerable effort has been focused on studying the zeroth order scalar curvature invariants in order to determine the minimal basis of algebraically independent invariants needed to generate any other scalar curvature invariant \cite{CM1991}. However, less is known about the higher order SPIs involving the covariant derivatives of the curvature tensor \cite{Exact,CMM2013}. 

Motivated by the results of \cite{musoke2016scalar} where the C.K. algorithm was employed to produce a minimal set of algebraically independent SPIs required for the 3D Szekeres cosmological spacetimes using a 3D analogue of the Newman Penrose spinor formalism \cite{MW2013}. In this paper, we will use the 3D G\"odel-like spacetimes as an example to argue that a set of SPIs chosen using the C.K. algorithm is sufficient to uniquely characterize a spacetime. In section \ref{sec1}, we introduce the quantities required for the coframe formalism in the G{\"o}del-like spacetimes. In section \ref{sec2}, we outline the C.K. algorithm and  present a lemma detailing the Segre types (or P-types in the notation of \cite{MW2013}) for the Ricci tensor with corresponding constraints on the metric functions. In section \ref{sec3}, for each P-type we present a summary of the ensuing iterations of the C.K. algorithm. In Section \ref{sec4}, we identify the subset of algebraically independent Cartan invariants and use them to generate a minimal set of SPIs that uniquely characterize each subcase of the G\"odel-like spacetimes. In section \ref{conclusion} we conclude the paper and discuss future work.
\end{section}

\begin{section}{The Coframe Formalism}	\label{sec1}
The 3D G\"{o}del-like metric is defined as \cite{Sousa}
\beq ds^2 = -[dt + H(r)d\phi]^2 + D^2(r)d\phi^2 + dr^2. \nonumber \eeq
\noindent Using the formalism in \cite{MW2013}, with \beq F_{\pm} = H \pm D, \label{Feqn} \eeq 
\noindent the coframe $\{\theta^a\}$ may be written in compact form:
\beq -n_{~\mu} =  &\theta^0_{~\mu} &= -\frac{1}{\sqrt{2}}(dt+F_{-}d\phi),\nonumber \\
m_{~\mu}= &\theta^1_{~\mu}& = \sqrt{2}dr,  \label{Nullframe} \\
-\ell_{~\mu} =  &\theta^2_{~\mu}& = -\frac{1}{\sqrt{2}}(dt+F_{+}d\phi).  \nonumber \eeq
\noindent Here, greek indices will be used for the coordinate basis and the indices $a,b,c...$ denote the (co)frame basis. We will denote the frame basis by $e_0^\mu = \ell^\mu, e_2^\mu = m^\mu$ and $e_2^\mu = n^\mu$. 


The connection one-forms, $\omega_{ab} = \Gamma_{acb} \theta^c$, which arise from the first Cartan structure equation and satisfy $\omega_{ab} =- \omega_{ba}$, are then:
\beq \omega_{01}  &=& \frac{F_{+}'}{2\sqrt{2}D}\theta^0 - \frac{D'}{2\sqrt{2}D}\theta^2, \nonumber \\
 \omega_{02} &=& \frac{H'}{2\sqrt{2}D} \theta^1,   \nonumber \\
 \omega_{12} &=&  \frac{D'}{2\sqrt{2}D} \theta^0 + \frac{F_{-}'}{2\sqrt{2}D}\theta^2,  \nonumber \eeq
%

\noindent where primes denote differentiation with respect to $r$. Using the second Cartan structure equation, we can compute the Ricci tensor, $R_{ab}$, and compute the quantities defined in \cite{MW2013} for the trace-free part of the Ricci tensor $ S_{ab} = R_{ab} - \frac{R}{3}g_{ab}$ we find that $\Psi_{1} = S_{10} = 0$, $\Psi_{3} = S_{12}= 0$ and the non-zero components are: 
\beq \begin{split} \Psi_{0} = S_{00} &= -\frac12 \left[ \frac{D''}{D} - \left(\frac{H'}{D} \right)^2 - \left(\frac{H'}{D}\right)' \right]   \\
\Psi_{2}  = S_{11} = S_{02} &= -\frac{1}{6} \left[ \frac{D''}{D} - \left(\frac{H'}{D}\right)^2 \right]     \\
\Psi_{4} = S_{22} &= -\frac12 \left[\frac{D''}{D} - \left(\frac{H'}{D} \right)^2 + \left(\frac{H'}{D}\right)' \right]  \\
R &=  -\frac12 \left[ \frac{4D''}{D} -  \left(\frac{H'}{D}\right)^2 \right].    \end{split} \label{MetricPhi} \eeq
\noindent To normalize the components of the Ricci tensor we must use the frame transformations to bring the components in agreement with table 1 of \cite{MW2013}. To summarize the frame freedoms we have a boost: 
\beq \tilde{\theta}^0 = \sqrt{A} \theta^0,~~\tilde{\theta}^1 = \theta^1,~~ \tilde{\theta}^2 = \frac{1}{\sqrt{A}} \theta^2, \label{boosteqn}  \eeq
\noindent and null rotations about $n^\mu$ and $\ell^\mu$ respectively:
\beq &\tilde{\theta}^0 = \theta^0,~~\tilde{\theta}^1= \theta^1 + B \theta^0,~~\tilde{\theta}^2 = \theta^2+ B \theta^1 + \frac{B^2}{2} \theta^0, &  \\
&\tilde{\theta}^0 = \theta^0+ C \theta^1 + \frac{C^2}{2} \theta^2,~~\tilde{\theta}^1= \theta^1 + C \theta^2,~~\tilde{\theta}^2 = \theta^2,&  \eeq
\noindent While we will work primarily with these Lorentz transformations, at times it will be helpful to express the null coframe as an orthonormal coframe $\{ t_\mu, x_\mu, m_\mu\}$ and rotate the spatial one-forms $x_\mu$ and $m_\mu$.
\end{section}

\begin{section}{The Zeroth iteration of the C.K. Algorithm: Segre types for the G{\"o}del-like Spacetimes} \label{sec2}

To put the Segre types into context, we reiterate the five steps of the Karlhede algorithm for the zeroth iteration, $q = 0$, for each P-type:
\begin{enumerate}
\item Calculate the set $\mathcal{R}_q$, the components of the derivatives of the curvature up to the q-th order. 
\item Fix the frame as much as possible by putting the elements of $\mathcal{R}_q$ into a canonical form.
\item Find the frame freedom given by the isotropy group $H_q$ of transformations which leave the canonical form (also known as normal form) of $\mathcal{R}_q$ unchanged.
\item Find the number $t_q$ of functionally independent functions of spacetime coordinates in the elements
of $\mathcal{R}_q$.
\item For $q>0$, if the isotropy group $H_q$ is the same as $H_{q - 1}$ and the number of functionally independent functions $t_q$ is equal to $t_{q-1}$ , then let $q = p + 1$ and stop. Otherwise set $q=q+1$ and repeat steps 1-5 until the algorithm stops.
\end{enumerate}

By studying the effect of the frame transformations  which {\bf do} change the form of the Ricci tensor, we may use the frame transformations to fix the coframe and determine all possible Segre types. For example, in the general case of the G{\"o}del-like spacetimes with Ricci tensor of Segre type [11,1], $SO(1,2)$ acting as the frame transformation group may be entirely fixed to put the tensor into the canonical form for P-type {\bf I} (or {\bf IZZ}) defined below. Contrast this with the opposite extreme, when the Ricci tensor is of  Segre type [(11,1)] and all of the frame transformations leaves the Ricci tensor unchanged. 

Noting that for the G{\"o}del-like spacetimes $\Psi_1 = \Psi_3 = 0$, we have the following distinct subclasses.

\begin{lem} \label{lem:Segre}
The G{\"o}del-like spacetimes admit the following Segre types for the Ricci tensor, which gives constraints on the metric functions $D(r)$ and $H(r)$ and a canonical form for the Ricci tensor: 

\begin{itemize}
\item $[11,1]$ (P- type {\bf I} or {\bf IZZ}) if  $D(r)$ and $H(r)$ satisfy
\beq \frac{D''}{D} - \left( \frac{H'}{D}\right)^2 = f, \nonumber \eeq
\noindent where $f$ is an arbitrary non-vanishing function with $f<0$ for P-type {\bf I} and $f>0$ for P-type {\bf IZZ}. 

\noindent Applying a boost \eqref{boosteqn}, the components of the Ricci spinor are:
\beq \tilde{\Psi}_{0} = \tilde{\Psi}_{4} = \sqrt{|\Psi_{0} \Psi_{4}|},~ \tilde{\Psi}_{2} = \Psi_{2},~ \tilde{R} = R, \nonumber \eeq
\noindent where the original components are defined in equation \eqref{MetricPhi}

\item $[1 z \bar{z}]$  (P- type {\bf IZ}) if for an arbitrary function $f(r)$ with $f'\neq 0$:
\beq \frac{H'}{D}  = f(r),~ \frac{D''}{D} - f^2 = 0.  \nonumber \eeq 
\noindent The components of the Ricci tensor are:
\beq \Psi_0 = - \Psi_4=\frac12 f' ,~~\Psi_2 = 0,~ R = -\frac32 f. \nonumber \eeq
\item $[12]$ (P-type {\bf IIZ}), if for an arbitrary function $f(r)$ with $f'\neq 0$:
\beq \frac{H'}{D} = f(r),~~ \frac{D''}{D} - f^2 + f' = 0. \nonumber \eeq
\noindent Applying a boost \eqref{boosteqn}, the components of the Ricci tensor are: 
\beq \tilde{\Psi}_{0} = 1,~\tilde{\Psi}_{4} = 0,~\tilde{\Psi}_{2} = \frac{f'}{6},~\tilde{R} = -\frac12 [3f^2 + 4f' ]. \nonumber \eeq 
\item $[(11),1]$ (P-type {\bf DZ}) if
\beq H' = C D,~~C \in \mathbb{R}. \nonumber \eeq
\noindent The components of the Ricci tensor are: 
\beq \Psi_{0} = \Psi_{4} = 3\Psi_{2} =  -\frac{1}{2} \left[ \frac{D''}{D} - C^2 \right],~R= -\frac12\left[ \frac{4D''}{D} -  C^2 \right].   \nonumber \eeq
\item $ [(11,1)]$ (P-type {\bf O}) if
\beq H' = C D,~ D'' - C^2D = 0,~C \in \mathbb{R}. \nonumber \eeq
\noindent The components of the Ricci tensor are: 
\beq \Psi_{0} = \Psi_{4} = 3\Psi_{2} =  0,~R = -\frac{3C^2}2. \nonumber \eeq
\end{itemize}  
\end{lem}

For each P-type of the Ricci tensor, we have a separate instance of the C.K. algorithm. We can use the components of the Ricci tensor in normal form to produce simpler invariants, and as they are all functions of $r$ we have {\it at most one functionally independent function} allowing us to identify $t_0$ for each case. The dimension of the isotropy group, $H_0$, of the Ricci tensor ($\mathcal{R}_0$) is found by counting the remaining frame freedom; for all P-types except P-type {\bf DZ} the dimension is ${\it dim}(H_0) =0$, while P-type {\bf DZ} has ${\it dim}(H_0) = 1$. To continue, we set $q=1$ and evaluate the covariant derivative of the Ricci tensor to compute $\mathcal{R}_1$.

In general, to distinguish a 3D G\"odel-like spacetime from another 3D spacetime, the C.K. algorithm requires the form of the Ricci tensor and its covariant derivatives up to appropriate order relative to a fixed coframe. However, we are interested in uniquely characterizing spacetimes within the G\"odel-like spacetimes, we will only focus on the algebraically independent invariants. For the class of G\"odel-like spacetimes, the form of these invariants will be sufficient to determine if two G\"odel-like spacetimes are distinct or not, since all other components of the Ricci tensor and its covariant derivatives will be expressed in terms of the algebraic independent invariants in a generic manner.

\end{section}

\begin{section}{The first iteration of the Cartan-Karlhede algorithm} \label{sec3}

The next iteration of the C.K. algorithm requires that we must compute the covariant derivative of the Ricci tensor, or equivalently covariant derivatives of the Ricci spinor, although we will focus on tensors due to the simplicity of the metric. It is here, where the frame derivatives of the Ricci tensor and the spin-coefficients are introduced as potential invariants. This can be seen by using the formula for the covariant derivative:
\beq R_{ab;c} = \nabla_c R_{ab}= R_{ab,c} - R_{db}\Gamma^d_{~ca} - R_{ad} \Gamma^d_{~cb}.
\label{CovRic} \eeq
\noindent We will repeat the analysis of the C.K. algorithm for those G{\"o}del-like spacetimes admitting a Ricci tensor with each P-type listed above. 

\begin{subsection}{P-type {\bf I} or {\bf IZZ}} \label{PtypeI}

According to Lemma \eqref{lem:Segre} at zeroth order the set of functions $\{ R, \Psi_{2}, \Psi_{0}\}$ are all non-constant invariants. With some algebra, these give rise to a simpler but algebraically equivalent set $\left\{ (\frac{H'}{D})^2,~~\frac{D''}{D},~~\left(\frac{H'}{D} \right)' \right\}$ ; the number of functionally independent invariants is $t_0 = 1$ as these are all functions of $r$ alone. We have fixed all of the isotropy at first order (${\it dim}(H_0) = 0$) and produced one functionally independent invariant and two essential {\it algebraically independent} classifying functions, \beq \{ I_0, I_1, I_2\} = \left\{ \left(\frac{H'}{D}\right)^2,~~\frac{D''}{D},~~\left(\frac{H'}{D} \right)' \right\}.\nonumber \eeq 

We proceed to the next iteration of the algorithm to show that the algorithm must terminate, since ${\it dim}(H_1) = 0$ and  $t_1 =1$, and more importantly, to collect all essential classifying functions.  As all isotropy has been fixed, the metric coframe is an invariant coframe, and so the frame derivatives of all zeroth order invariants are now invariants as well. Using this fact we may solve for the frame derivatives of the Ricci tensor components $\Psi_0, \Psi_2$ and $R$ and the spin-coefficients \cite{MW2013} from the components of the covariant derivative of the Ricci tensor: 

\beq & \kappa = \frac{A F_{+}'}{2\sqrt{2}D},~\sigma=0,~\tau = \frac{D'}{2\sqrt{2}D},& \nonumber \\
&\epsilon = 0,~2\alpha = \frac{H'}{2\sqrt{2}D} + \frac{e_1(A)}{A},~\gamma = 0,& \label{spincoefs}\\
&\pi = -\frac{D'}{2\sqrt{2}D},~~\lambda = 0,~\nu = \frac{F_{-}'}{2\sqrt{2} A D},& \nonumber \eeq

\noindent where the boost parameter in \eqref{boosteqn} is defined as $A = \sqrt{\frac{\Psi_{0}}{\Psi_{4}}}$. 

There are four algebraically independent invariants appearing at first order. From the spin-coefficients we find $2\sqrt{2}\tau = (ln D)'$ and $\alpha$ from which we can solve for the sign of $H'/D$. Taking linear combinations of the frame derivatives of $\Psi_2$ and $R$ we have two algebraically independent invariants, namely $e_2( I_1)$ and $e_2(I_2)$. Thus the list of algebraically independent invariants required to classify the spacetime is 

{\small \beq &  \{ I_0, I_1, I_2; I_3,I_4,I_5, I_6 \} = \left\{ \left(\frac{H'}{D}\right)^2,\frac{D''}{D},\left(\frac{H'}{D}\right)'; \frac{D'}{D}, \left(\frac{D''}{D}\right)', \left(\frac{H'}{D}\right)'', sign\left(\frac{H'}{D}\right)  \right\}.&  \label{templateI} \eeq }

\end{subsection}

\begin{subsection}{P-type {\bf IZ}}
According to Lemma \eqref{lem:Segre}, at zeroth order the set of functions $\{ R, \Psi_{0}\}$ are the only non-constant invariants, which give rise to the simpler set of invariants: 
\beq \{ I_0, I_1, I_2 \} = \left \{ f^2, f^2, f' \right \}. \nonumber \eeq
%
\noindent  The number of functionally independent invariants is $t_0 = 1$ as these are all functions of $r$ alone. We have fixed all of the isotropy at first order (${\it dim}(H_0) = 0$) and produced one functionally independent invariant and one algebraically independent essential classifying function. 

The algorithm stops at first order since ${\it dim}(H_1) = 0$ and $t_1 =1$. The metric coframe is an invariant coframe, and so we may separate the spin-coefficients and the frame derivatives of the Ricci tensor components \cite{MW2013} from the components of the covariant derivative of the Ricci tensor. Comparing with \eqref{templateI}, the number of algebraically independent invariants has been reduced: 


{\small \beq \{ I_0, I_1, I_2; I_3, I_4, I_5, I_6 \} = \left\{ f^2,f^2,f'; \frac{D'}{D}, 2f'f,  f'', sign(f)  \right\}. \label{IZtemplate} \eeq}

\end{subsection}

\begin{subsection}{P-type {\bf IIZ}}
From Lemma \eqref{lem:Segre}, at zeroth order the set of functions $\{ R, \Psi_{2}\}$ are non-constant invariants, from which we have the following set of invariants $\left\{f^2,~f'\right\}$. The number of functionally independent invariants is $t_0 = 1$, and  we have fixed all of the isotropy at zeroth order, yielding the following set of invariants 
\beq \{ I_0, I_1, I_2 \} = \left\{ f^2,~f^2-f', f' \right\}.\nonumber \eeq 

Yet again, the frame is an invariant coframe and we can isolate the invariants in \eqref{templateI} from the components of the covariant derivative of the Ricci tensor to completely classify this spacetime:

{\small \beq \{ I_0, I_1, I_2; I_3, I_4, I_5, I_6 \} = \left\{f^2,f^2-f', f'; \frac{D'}{D}, 2f'f-f'',  f'', sign(f)  \right\}. \label{IIZtemplate} \eeq} 

\noindent As in the previous case, the number of algebraically independent Cartan invariants has been reduced. 
\end{subsection}

\begin{subsection}{P-type {\bf DZ}}
Lemma \eqref{lem:Segre} indicates that at zeroth order the set of functions $\{ R, \Psi_{2}\}$ are the only distinct non-constant invariants. The number of functionally independent invariants is $t_0 = 1$. Thus, we have fixed most of the isotropy at first order except spatial rotations, i.e., ${\it dim}(H_0) = 1$, and the reduced Cartan invariants are \beq \{ I_0, I_1, I_2\} = \left\{ C^2,\frac{D''}{D},0 \right\}.\nonumber \eeq 

At the first iteration the metric coframe is not yet an invariant coframe, and so we must be careful with taking frame derivatives of invariants until an invariant coframe is determined. To entirely fix the frame,  we consider the transformation rules for the spin-coefficients under an element of $SO(2)$:

\beq (\ell+n)' = LHS,~~(\ell-n \pm 2im)' = e^{\mp i\tilde{t}}(\ell-n \pm 2im). \nonumber\eeq

\noindent This produces the following transformation rules for the spin-coefficients and curvature components \cite{MW2013}:

\beq &(\gamma + \sigma - \epsilon - \lambda)' = LHS,& \nonumber \\ 
&(4\alpha+\kappa-\pi+\nu-\tau)' = LHS,& \nonumber \\
&(2(\gamma + \epsilon) \pm i(\kappa-\pi + \tau - \nu))' = e^{\pm i\tilde{t}} LHS, & \nonumber \\
&(4\alpha + \pi - \kappa + \tau - \nu + \pm 2i(\epsilon-\gamma+\sigma-\lambda))' = e^{\pm 2 i\tilde{t}} LHS,& \nonumber \\
&(\lambda + \sigma - \gamma - \epsilon + \pm i(\pi - \tau))' = e^{\pm i\tilde{t}}(LHS - \delta \tilde{t} \mp \frac{i}{2} (D\tilde{t} - \Delta \tilde{t})),& \nonumber \\
&(\kappa + \pi + \nu + \tau)' = LHS-(D\tilde{t} + \Delta \tilde{t}),& \nonumber \\
&(\Psi_0 + 2\Psi_2 + \Psi_4)' = LHS, & \nonumber \\
&(\Psi_0 - \Psi_4 \pm 2i(\Psi_1 + \Psi_3))' = e^{\mp i\tilde{t}} LHS, \nonumber \\
&(|Psi_0 - 6\Psi_2 + \Psi_4 \pm 4i (\Psi_1 - \Psi_3))' = e^{\mp 2 i\tilde{t}} LHS.& \nonumber \eeq

\noindent Substituting the non-zero spin-coefficients and curvature components \eqref{spincoefs} we find the simpler transformation rules for the G{\"o}del-like spacetimes:

\beq & 2\gamma ' =0,~ \sigma' - \lambda' = 0,~ 8\alpha' = 8 \alpha,~ \nu' = 2\alpha' + \tau',~\kappa' = 2\alpha - \tau', & \nonumber \\
&2[\pi' + \tau' ] = -D\tilde{t} - \Delta \tilde{t},~2\sigma' \pm i(\pi' - \tau') = e^{\pm i \tilde{t}}[ - \delta \tilde{t} \mp i(2\tau-\frac{D \tilde{t}}{2} + \frac{\Delta \tilde{t}}{2})] & \nonumber \eeq

Comparing with the spin-coefficients of original frame (with $A=1$):  $\tau = - \pi$,  $\kappa = 2\alpha+\tau$, and $\nu = 2\alpha-\tau$; we conclude that setting the rotation parameter $\tilde{t}$ to zero is the best choice.  Any other choice of $\tilde{t} \neq 0$ would cause $\sigma \neq 0$, $\tau \neq \pi$, and the frame derivatives of any scalar with respect to $e_0$ and $e_2$ to be non-zero; effectively increasing the number of invariants instead of decreasing the number.

Fixing $\tilde{t}=0$, and simplifying the components of covariant derivative of the Ricci tensor, one finds the first order invariants are $e_1( \Psi_2), e_1( R)$, and the spin coefficients. Of the spin-coefficients, the only algebraically independent invariant appears in $\alpha$, since $\tau = -\pi$ does not appear in the first covariant derivative of the Ricci tensor. Noting that $C - 4\sqrt{2}\alpha = 0$, $\alpha$ provides a useful discrete invariant as the sign of the constant C. Returning to the frame derivatives, we have one algebraically independent invariant appearing: $e_1( I_2)$. Thus the list of invariants required to classify the spacetime is:

\beq \{I_0, I_1, I_2; I_4, I_5, I_6 \}  = \left\{ C^2, \frac{D''}{D}, 0; \left(\frac{D''}{D}\right)', 0, sign(C)  \right\}.\nonumber \eeq

%

\begin{subsubsection}{The second iteration of the Cartan-Karlhede algorithm: P-type {\bf DZ}}

Applying the frame derivatives to the first order invariants, we produce two new algebraically independent classifying function:

{\small \beq \begin{split}  \{I_0, I_1, I_2; & I_4, I_5, I_6; I_3I_4, I_7 \}  = \\ & \left\{ C^2, \frac{D''}{D}, 0; \left(\frac{D''}{D}\right)', 0, sign(C) ; \frac{D'}{D}\left(\frac{D''}{D}\right)', \left(\frac{D''}{D}\right)''  \right\}.\end{split} \eeq}

\noindent We note that, if $I_4 \neq 0$ then we can isolate $I_3$. However, for the P-type {\bf DZ} spacetimes, if $I_4=0$ then all appearances of $I_3$ vanish within the components of the second covariant derivative of the Ricci tensor.



\end{subsubsection}

\end{subsection}

\begin{subsection}{P-type {\bf O}}
As all components of the Ricci tensor are constant, using Lemma \eqref{lem:Segre} and the form of the spin-coefficients \eqref{spincoefs}, one may show that the covariant derivative of the Ricci tensor is zero. We conclude that these are locally homogeneous spaces which are fully determined at zeroth order by the constant $C^2$.
%
%
\noindent In fact, the P-type {\bf O} G\"odel-like spacetimes are all isometric to the 3D anti-de Sitter spacetime.
\end{subsection}

\end{section}

\begin{section}{Characterization of the G\"odel-like spacetimes using Scalar polynomial invariants} \label{sec4}

To begin, we consider the Cartan invariants of the P-type {\bf I} (or {\bf IZZ}) 3D G\"{o}del-like metric and compare them to the SPIs given in \cite{CMM2013}. In section \ref{PtypeI} we found that the algebraically independent invariants required for classification of the P-type {\bf I} metrics are 
\beq \begin{split}   \{ I_0, &I_1, I_2; I_3, I_4,I_5, I_6 \} = \\ & \left\{ \left(\frac{H'}{D}\right)^2,\frac{D''}{D},\left(\frac{H'}{D}\right)'; \frac{D'}{D}, \left(\frac{D''}{D}\right)', \left(\frac{H'}{D}\right)'', sign\left(\frac{H'}{D}\right)  \right\}.\end{split} \eeq
	Calculating the SPIs and making the substitutions for the zeroth order Cartan invariants:
{\small	\begin{eqnarray} 
		\frac{H'}{D} &=& I_6 \sqrt{I_0},
		\\
		\frac{D''}{D} &=& I_1,
		\\
		\frac{H''}{D} &=&
		\frac{H''}{D} - \left( \frac{H'}{D} \right)' + \left( \frac{H'}{D} \right)'
		\\
		\nonumber
		&=& \frac{H''}{D} - \left( \frac{H''}{D} - \frac{H'}{D} \frac{D'}{D} \right) + \left( \frac{H'}{D} \right)'
		= \sqrt{I_0} I_6 I_3 + I_2.
	\end{eqnarray}
	The three zeroth order SPIs in \cite{CMM2013} are:
	\begin{eqnarray}
		R_0 &=& R^a_{~a} =
		-1/2\,{\frac {4\, \left( D \right) {\it \Drr}-{{\it \Hr}}^{2}}{{
D}^{2}}}
=
-2\,I_1+1/2\,I_0,
		\label{SPI1}\\
		R_1 &=& R^{ab} R_{ab}
		=
2\,{I_1}^{2}-2\,I_1I_0+3/4\,{I_0}^{2}-
1/2\,{I_2}^{2},
		\label{SPI2} \\
		R_2 &=& R^{ab} R_a^{~c} R_{bc}
		=
-2\,{I_1}^{3}+3\,{I_1}^{2}I_0-3/2\,I_1{{\it 
\Izero}}^{2}+3/4\,I_1{I_2}^{2}+1/8\,{I_0}^{3}, \label{SPI3}	
	\end{eqnarray}}
	which are polynomials in $\sqrt{I_0}, I_1, I_2$. 
	
	At the next order, we make the similar substitutions
	to express nine of the first order SPIs in terms of the Cartan invariants. Here we give only five; all three of the case $[1,1]$ and two of the case $[1,1,1,1]$ in the categorization of \cite{CMM2013}:
{\small	\begin{align}
		R_3 &=R^{;a} R_{;a}
		\;
		=
\left( \sqrt {I_0} I_6I_2-2\,I_4 \right) ^{2}, \label{SPI4}
 \end{align} \begin{align}
		R_4 = R^{bc;d} R_{bc;d}
		&=
-{I_0}^{3}+2\,{I_4}^{2}-1/2\,{I_5}^{2}-{I_1}^{
2}I_0+2\,I_1{I_0}^{2}+I_1\sqrt {I_0}I_5 \label{SPI5}
\\ \nonumber
&
-I_3{I_0}^{3/2}I_2-5\,\sqrt {I_0}I_2I_4+I_1I_3\sqrt {I_0} I_6I_2-{I_0}^{3/2}I_5 \nonumber 
\\ \nonumber
&
+{\frac {23}{4}}\,I_0\,
{I_2}^{2}-1/2\,{I_3}^{2}{I_2}^{2},
\end{align} \begin{align} 
		R_5 = R^{bc;d} R_{bd;c}
		&=
1/2\,{I_0}^{3/2}I_3I_2-1/2\,I_1I_3\sqrt {I_0}I_2+{\frac {9}{8}}\,I_0\,{I_2}^{2}+
{I_4}^{2} \label{SPI6}
\\ \nonumber
&
+1/2\,{I_0}^{3}+1/2\,{I_1}^{2}I_0-I_1{I_0}^{2}-1/2\,I_1\sqrt {I_0}I_5 
\\ \nonumber
&
-3/2\,\sqrt {I_0} I_6I_2I_4+1/2\,F_{
{2}}I_5I_2+1/2\,{I_0}^{3/2}I_5 ,
\end{align} \begin{align}
		R_6 &=R^{;c} {R_c}^{e;f} R_{;e} R_{;f}
		\;
		=
 \left( \sqrt {I_0} I_6I_2-2\,I_4 \right) ^{3} \left( 
\sqrt {I_0} I_6I_2-I_4 \right),  \label{SPI7}
\end{align} \begin{align}
		R_7 &= R^{;a} R_a^{~b;c} R_{c~;b}^{~d} R_{;d}
		\;
		=
\frac{1}{8}\, \left( \sqrt {I_0} I_6I_2-2\,I_4 \right) ^{2}
 \left( 9\,I_0\,{I_2}^{2}-16\,\sqrt {I_0} I_6I_2 I_4+8\,{I_4}^{2} \right).  \label{SPI8}
	\end{align}}
\noindent The other SPIs give more expressions of the same form: they are polynomials in $\left\{ \sqrt{I_0}, I_1, I_2 ; I_3, I_4, I_5, I_6 \right\}$.  
	
	While we have only considered P-type {\bf I}, the SPIs we have listed above may be adapted to the G\"odel-like spacetimes of P-type {\bf IZ}, and {\bf IIZ}. These two P-type is can be identified by the following syzygies between the zeroth order SPIs, which are unique to their P-type:
\vspace{ 3 mm}

\noindent P-type {\bf IZ}
\beq R_1 R_0 + R_2 = \frac{16}{9} R_0; \eeq
\noindent P-type {\bf IIZ}
\beq R_2 - R_1 R_0 + \frac29 R_0^3 = \frac{1}{36} (6 R_1 + 2 R_0^2)^{\frac32}. \eeq
\noindent The reduction in the number of algebraically independent Cartan invariants changing from P-type {\bf I} to {\bf IZ} or {\bf IIZ} is also reflected in the set of algebraically independent SPIs.

In the case of P-type {\bf DZ} and {\bf O}, the different structure of the algebraically independent Cartan invariants:
{\small \beq \begin{split}  \{I_0, I_1, I_2; & I_4, I_5, I_6; I_3I_4, I_7 \}  = \\ & \left\{ C^2, \frac{D''}{D}, 0; \left(\frac{D''}{D}\right)', 0, sign(C) ; \frac{D'}{D}\left(\frac{D''}{D}\right)', \left(\frac{D''}{D}\right)''  \right\},\nonumber \end{split} \eeq}

\noindent is reflected in the set of SPIs. Since the Cartan invariants $I_2$ and $I_5$ vanish, this eliminates $I_3$ from the set of SPIs \eqref{SPI1}- \eqref{SPI3} and \eqref{SPI4}-\eqref{SPI8}.  While P-type {\bf O} is easily identified from the SPIs, P-type {\bf DZ} can be determined by checking the following two syzygies for the first order SPIs: 

\beq R_4^2 = 2 R_6, \text{ and } R_4^2 = 4 R_7 \eeq 

\noindent In this case we must introduce two second order SPIs to  recover the Cartan invariants $I_3$ and $I_7$. Two possibilities are: 
\beq R_8 = R_{3;a} R^{3;a} = \sqrt{2} 4 I_4 I_7 \eeq
\noindent and  
 \beq \begin{split} R_9 &= S^{ab} S^{cd} S_{ab;cd} = \frac13 I_0^4 - I_0^3 I_1 - \frac29 I_0^2 I_7 -\frac29 I_0^2 I_3 I_4 + I_0^2 I_1^2 \\ & +\frac49 I_0 I_1 I_7 + \frac49 I_0 I_1 I_3 I_4 - \frac13 I_1^3 I_0 - \frac29 I_1^2 I_7- \frac29 I_1^2 I_3 I_4. \end{split} \eeq
\noindent Including these two SPIs we now have a set containing all of the algebraically independent Cartan invariants. This set will characterize the P-type {\bf DZ} spacetimes.

	We have found that a representative sample of the SPIs can be expressed as polynomials in (the square root of) the Cartan invariants. 
As any combination of invariants is an invariant, this suggests that an appropriate choice of the above ten SPIs may be sufficient to classify all G\"odel-like metrics, as long as one is concerned with functional dependence. 

\end{section}

\begin{section}{Conclusion} \label{conclusion}

In this paper we have applied the Cartan-Karlhede algorithm to the 3D G\"{o}del-like spacetimes to generate a sequence of algebraically independent Cartan invariants that fully characterize each subclass of these spacetimes. Beyond the ability of the Cartan invariants to classify all spacetimes, including the locally homogeneous spacetimes and the degenerate Kundt spacetimes \cite{CSI1, CSI2, CSI3, Kundt2009}, the invariants arising from the C.K. algorithm provide a helpful framework to study SPIs for spacetimes which are characterized by $\mathcal{I}$. While the invariants in each sequence can only be considered as an invariant relative to the choice of a particular coframe; in theory, working with such a coframe is not a hindrance \cite{MacCallum2015}. However, in some applications choosing this frame may not be possible, such as in higher dimensions and in numerical GR. In these circumstances, SPIs are better suited as invariants. 

As the G\"odel-like spacetimes are characterized by $\mathcal{I}$, we have examined the set of SPIs formed from the Ricci tensor, and its covariant derivatives up to second order. By comparing with the results of the C.K. algorithm we have identified a minimal subset of SPIs which uniquely characterize these spacetimes, and identify the various subcases relative to P-type in terms of additional syzygies between the SPIs. This work is meant to be a continuation of \cite{musoke2016scalar}, where a minimal basis of algebraically independent SPIs were determined for the Szekeres cosmological spacetimes using the C.K. algorithm. Unlike the G\"odel-like spacetimes, the Szekeres spacetime does not contain subcases of differing P-type and did not illustrate the ability of the chosen SPIs to distinguish P-type. 

	We believe this approach will be helpful for identifying a minimal set of SPIs (up to the appropriate order) needed to uniquely characterize a spacetime. While the Newman-Penrose formalism is already employed in 4D for the zeroth order invariants \cite{CM1991}, the C.K. algorithm allows for a reduction of the number of non-zero independent components of the covariant derivatives of the curvature tensor. It has been shown that partial knowledge of the Cartan invariants allows for the manipulation of first order SPIs \cite{gang,PageMcNutt, mcnutt2017}, this suggests we can potentially  simplify the form of any higher order SPIs using the C.K. algorithm. In future work we will explore the implications of the C.K. algorithm for general spacetimes of fixed P-type using the 3D spinor formalism, and extend this procedure to higher dimensions \cite{gang}.

\end{section}

\section*{Acknowledgements}

This work was supported in part through the Research Council of Norway, Toppforsk grant no. 250367: Pseudo-Riemannian Geometry and Polynomial Curvature Invariants: Classification, Characterisation and Applications (D.M.) and by the Natural Sciences and Engineering Research Council of Canada (N.K.M). Research at Perimeter Institute is supported by the Government of Canada through Industry Canada and by the Province of Ontario through the Ministry of Research and Innovation.

\bibliographystyle{unsrt-phys}

\bibliography{bibfile}

\end{document}